\documentclass{mn2e}
\usepackage{epsfig}

\newcommand{\gtsimeq}{\raisebox{-0.6ex}{$\,\stackrel
        {\raisebox{-.2ex}{$\textstyle >$}}{\sim}\,$}}

\newcommand{\mdot}{\dot{M}}
\newcommand{\msun}{M_{\odot}}
\newcommand{\msunyr}{\msun\,{\rm yr}^{-1}}

\title[A New Evolutionary Channel for Type Ia Supernovae]
{A New Evolutionary Channel for Type Ia Supernovae}
\author[A.R.~King, D.J.~Rolfe, K.~Schenker]{
A.R.~King, D.J.~Rolfe, K.~Schenker\\
Theoretical Astrophysics Group, University of Leicester,
Leicester, LE1~7RH, UK}

\begin{document}

\maketitle

\begin{abstract}
  We show that long--period dwarf novae offer a promising route for
  making Type Ia supernovae. For typical dwarf nova duty cycles $d
  \sim 0.1 - 0.01$, mass is accreted by the white dwarf mainly during
  dwarf nova outbursts at rates allowing steady nuclear burning of
  most of the accreted matter. Mass gains up to $\sim 0.4\msun$ are
  possible in this way. Although these are too small to allow a
  $0.7\msun$ WD to reach the Chandrasekhar mass, they are sufficient
  if the WD grew to $\ga 1\msun$ in a previous episode of
  thermal--timescale mass transfer, i.e. for those long--period dwarf
  novae which descend from supersoft binaries. A further advantage of
  this picture is that the supernova always occurs in a binary of
  small secondary/primary mass ratio, with the secondary having very
  little remaining hydrogen. Both features greatly reduce the
  possibility of hydrogen contamination of the supernova ejecta.
\end{abstract}


\section{Introduction}

It is now generally agreed that Type Ia supernovae originate from
accreting white dwarfs, and widely accepted that their occurrence
signals arrival at the Chandrasekhar mass $M_{\rm Ch} \simeq
1.4\msun$. However there is no consensus as to how the white dwarf
gains mass (see e.g. Livio, 2001, for a review). Two possibilities are
currently discussed. In the first, two white dwarfs merge as their
relative orbit shrinks under gravitational radiation (the
double--degenerate scenario). In the second possibility, a white
dwarf accretes from a non--degenerate companion (the
single--degenerate scenario).  Observational evidence is currently too
sparse to give a strong preference to either picture. However both
have serious theoretical drawbacks.

For white dwarf mergers, many authors (Saio \& Nomoto, 1985, 1998;
Kawai, Saio \& Nomoto, 1987; Mochkovitch \& Livio, 1990; Timmes,
Woosley \& Taam, 1994; Mochkovitch, Guerrero \& Segretain, 1997) argue
that no explosion takes place and that the result is instead a quiet,
accretion--induced collapse (AIC), forming a neutron star.

The single--degenerate picture suffers from two main problems. First,
growth of the white dwarf mass $M_1$ requires efficient nuclear burning
of most of the accreted hydrogen. This happens only for accretion rates $\dot
M$ satisfying
\begin{equation}
\dot M_l \la {\dot M\over M_1/\msun-0.52} \la \dot M_h \simeq 2.5\dot M_l
\label{burn}
\end{equation}
where $\dot M_l \simeq 3.4\times 10^{-7}\,\msun {\rm yr}^{-1}$ (Nomoto
et al., 1979; Fujimoto, 1982). For lower $\dot M$ the burning occurs
in nova explosions. These expel matter and leave the white dwarf mass
essentially unchanged. For $\dot M/(M_1/\msun-0.52) > \dot M_h \simeq
8.5\times 10^{-7}\,\msun {\rm yr}^{-1}$ not all of the hydrogen is
burnt. In some pictures much of the excess is expelled in a
radiatively driven wind: in e.g. the strong wind solution of Hachisu
et al. (1996) the burning rate is limited by $\dot M_h$.  This allows
mass growth to continue, but for $\dot M >> \dot M_h$ this is clearly
an inefficient process, as only a small fraction of the transferred
mass is gained by the white dwarf. If there is no significant wind
things may be still worse, as the excess accretion will probably form
a common envelope around the binary, causing it to merge and become a
massive red giant. The second difficulty with this single--degenerate
scenario results from the proximity of a large H--rich companion at
the time the supernova explodes. This may contaminate the supernova
ejecta (Marietta, Burrows \& Fryxell, 2000), in conflict with the
defining characteristic of Type Ia SNe as having no detectable
hydrogen (see, however, Lentz et al. 2002).

The $\dot M$ constraint rules out most candidate progenitor binaries
for the single--degenerate scenario. Cataclysmic variables (CVs)
generally have mass transfer rates $-\dot M_2 << \dot M_l$ and so are
prone to losing the transferred mass in nova explosions. Supersoft
X--ray binaries (SSS) are more promising. Here in contrast to CVs the
companion/WD mass ratio $q = M_2/M_1$ is $\ga 1$, and mass transfer
shrinks the companion's Roche lobe relative to the stellar surface.
For a companion with a radiative envelope the result is mass transfer
on a thermal timescale, i.e. $\dot M = -\dot M_2 \sim M_2/t_{\rm KH}$.
This is close to the required range if the companion has mass $M_2
\sim 1\msun$ and thus a Kelvin--Helmholtz timescale $t_{\rm KH} \sim
3\times 10^7$~yr (van den Heuvel et al., 1992). The white dwarf in
such SSS binaries can gain mass. However, systematic calculations
(e.g. Langer et al., 2000) show that it may be difficult to grow $M_1$
from a typical value $0.7\msun$ all the way to $M_{\rm Ch}$. Thus only
systems starting with rather more massive white dwarfs give SNe Ia.
This in turn means that the main--sequence progenitor of the white
dwarf must itself have been rather massive, significantly reducing the
number of systems which can give SNe Ia by this channel. In addition,
this type of evolution is vulnerable to the hydrogen contamination
constraint. Not only is the companion fairly massive and
hydrogen--rich, its Roche lobe offers a large target for the expanding
SN shell as the mass ratio $q$ must be $\ga 1$ when the explosion
occurs.

In this paper we reconsider the single--degenerate scenario, and offer
an alternative that avoids these difficulties.

\section{Long Period Dwarf Novae}

We have seen that SSS binaries do not readily produce SNe Ia because
the white dwarf mass does not grow enough. The accretion rate is
sensitive to the mass ratio $q$, and moves out of the narrow band
(\ref{burn}) in which efficient mass gain occurs before much mass is
transferred. An accretion rate which remains almost constant as $q$
decreases is clearly a major advantage for a plausible progenitor WD
binary class.

\begin{figure}
  \includegraphics[width=\columnwidth]{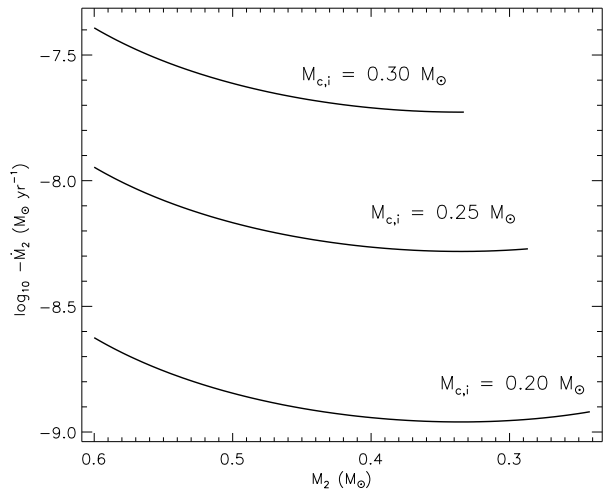} 
\caption{ The mass
  transfer rate $-\dot M_2$ given by Eq (36) of Ritter (1999) plotted
  as a function of $M_2$ for initial WD and donor masses $M_{\rm
  1,i} = 1\msun, M_{\rm
  2,i} = 0.6\msun$ and various initial donor core masses $M_{\rm c, i}$.}
  \label{Fig:Mdot}
\end{figure}

Only one type of WD binary satisfies this requirement: long--period
($\ga 2$~d) systems where the WD accretes from a low--mass red giant.
These systems are driven by the nuclear expansion of the red giant. The
helium core mass $M_c$ of this star increases slightly through
shell--burning during the binary evolution, causing considerable
envelope expansion. The envelope mass is depleted by shell--burning
but much more by mass transfer to the white dwarf. Calculations
(Webbink, Rappaport \& Savonije, 1983; Ritter, 1999) show that the
mass transfer rate varies remarkably little while most of the
companion's envelope is transferred, the value of the rate being fixed
by the core mass or equivalently the binary period. Eq (36) of Ritter
(1999) relates $-\dot M_2$ directly to the companion mass $M_2$ for
the case where all the transferred mass is accreted by the white
dwarf. Fig. \ref{Fig:Mdot}
%%%%%%%%%%%%%%%%%%%%%%%%%%%%%%%%%%%%%
%%%%%%%%%%%%%%%%%%%%%%%%%%%%%%%%%%%%%
%%%%%%%%%%%%%%%%%%%%%%%%%%%%%%%%%%%%%
\begin{figure}
%*
\includegraphics[width=\columnwidth]{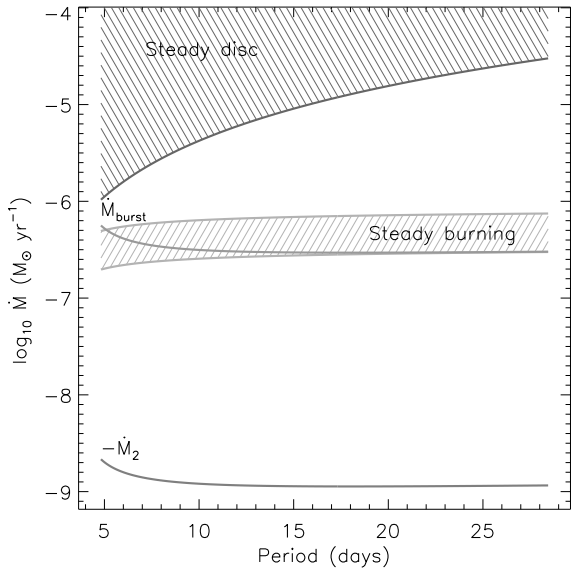}
  \caption{
    Binary evolution with mass transfer driven by nuclear evolution of
    the donor on the red giant branch (RGB). All mass lost from the
    donor is retained by the white dwarf. The mass transfer rate
    $-\mdot_2$ must avoid the upper hatched region for the disc to be
    unstable to dwarf nova outbursts.  The narrow hatched band is the
    efficient accretion region (\ref{burn}).  The assumed duty cycle
    is 0.004 and the masses of donor, core and white dwarf at the
    start of this phase are 0.53$\msun$, 0.2$\msun$ and 1.1$\msun$
    respectively.}
\label{Fig:rgbtracks1}
\end{figure}
%*
%%%%%%%%%%%%%%%%%%%%%%%%%%%%%%%%%%%%%
%%%%%%%%%%%%%%%%%%%%%%%%%%%%%%%%%%%%%
\begin{figure}
%*
\includegraphics[width=\columnwidth]{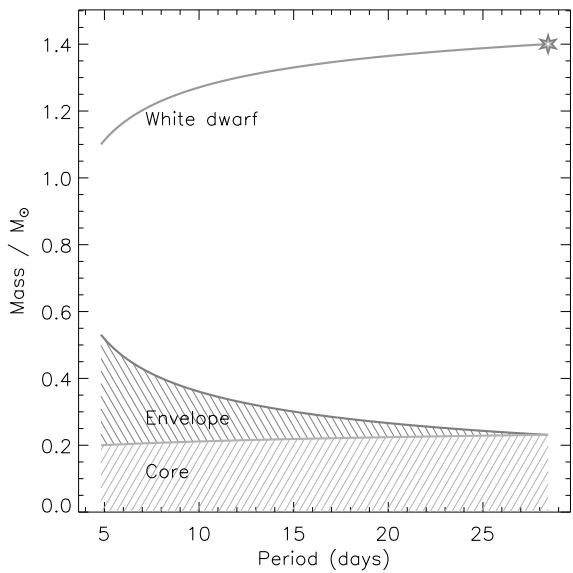}
  \caption{As for Figure \ref{Fig:rgbtracks1}.
%\emph{Right panel:} 
Evolution of the
    donor core and envelope masses, and the white dwarf mass. After
    $2.3\times10^8$ years the white dwarf reaches the Chandrasekhar mass and
    a Type Ia supernova ensues.}
  \label{Fig:rgbtracks2}
\end{figure}
%*
%%%%%%%%%%%%%%%%%%%%%%%%%%%%%%%%%%%%%
shows that for a WD starting at $M_{1,i} = 1\msun$ the mass transfer
rate $-\dot M_2$ changes by no more than factors $\sim 2$ as $M_2$
decreases from $M_{2,i} \sim 0.6\msun$ to $\sim 0.3\msun$.

While this is encouraging for a candidate SNe Ia progenitor there
appears at first sight to be a major difficulty. The mass transfer
rates in such binaries are typically only $\sim 10^{-8}\,\msun {\rm
  yr}^{-1}$, more than a factor 10 short of the steady burning regime
(\ref{burn}). However we must distinguish the mass {\it transfer} rate
$-\dot M_2$ from the mass {\it accretion} rate $\dot M$. All these
long--period systems are so wide that their accretion discs have
ionization zones; they are dwarf novae (King et al., 1997). This point
was previously neglected but is crucial for modeling these
long--period systems correctly. It is probably the shortness of the
duty cycle $d$, together with the low space density resulting from
their relatively rapid evolution that makes discovery of long--period
systems of this type difficult. Accretion on to the white dwarf occurs
almost entirely during outbursts, with accretion rates $\sim -\dot
M_2/d$ which can therefore be close to the efficient burning regime.

The mean duty cycle $d$ is presumably itself a function of the
evolutionary state of the binary, although the lack of understanding
of disc viscosity in current theory does not allow one to predict it
with any certainty. However for typical values $\sim 0.1$ to a few times
$10^{-3}$ one can always find initial white dwarf, donor and core
masses which keep the accretion rate during outbursts inside the
efficient burning regime (\ref{burn}) through most of the
evolution. This kind of mass gain therefore works even if wind loss is
inefficient for $\dot M > \dot M_h$. Figs. \ref{Fig:rgbtracks1},
\ref{Fig:rgbtracks2} show a case where $M_1$ is able to grow to
$M_{\rm Ch}$.
% for different assumed
%values of $d$ and initial masses $M_{1i}, M_{2i}$. With typical values
%$d \sim 0.1 - 0.01$ we see that there is now a distinct possibility
%that accretion will occur at rates allowing efficient mass gain. Note
%that the accretion rate during outbursts is always inside or very
%close to the efficient burning regime (\ref{burn}). This kind of mass
%gain therefore works even if wind loss is inefficient for $\dot M >
%\dot M_h$.

A further advantage for this picture is that the SN Ia explosion must
always occur for mass ratios $q < 1$, and indeed generally with very
little of the hydrogen envelope of the secondary remaining. This then
avoids the second difficulty (hydrogen contamination) afflicting the
single--degenerate scenario. The secondary is now a rather smaller
target for the expanding supernova shell, and in any case contains
little hydrogen.

\section{SN Ia Progenitors}

\begin{figure}
  \includegraphics[width=\columnwidth]{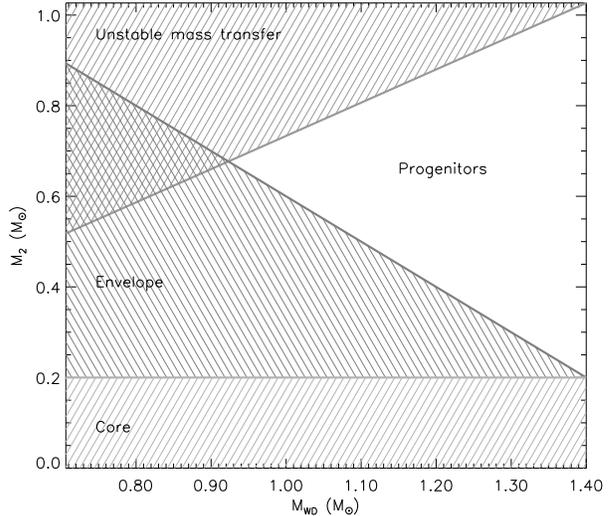} 
\caption{
  Component masses of potential SN Ia progenitors at the start of the
  conservative mass transfer phase on the RGB. Systems in the upper
  hatched region undergo dynamically unstable mass transfer and maybe
  a common envelope phase, rather than the evolutionary scenario
  considered here. Systems in the lower hatched region have
  insufficient mass in the donor envelope to raise the white dwarf to
  the Chandrasekhar limit. The upper limit on the white dwarf is the
  Chandrasekhar mass. Potential SN Ia progenitors therefore lie in the
  clear triangle region, with a minimum white dwarf mass $\sim1\msun$.
  The precise value depends on the core mass at this stage, which
  must be in the approximate range 0.2--0.3$\msun$, and the details of
  mass loss for $\dot M > \dot M_h(M_1/\msun-0.52)$.}
\label{Fig:progenitors}
\end{figure}

\begin{figure*}
  \includegraphics[width=0.7\textwidth]{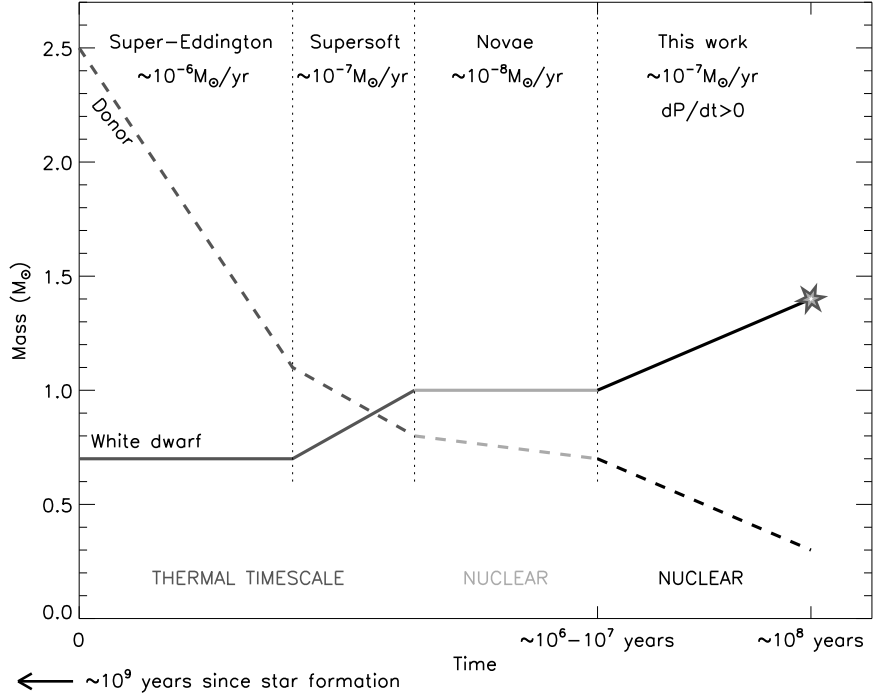} 
\caption{ The
  two--stage process proposed for progenitors of SNe Ia. A
  significantly nuclear--evolved $\sim2.5\msun$ star transfers mass to
  a $\sim0.7\msun$ white dwarf.  The high mass ratio $q$ leads to mass
  transfer on the timescale of the donor's thermal reaction to mass
  loss. In some cases there is an initial phase of non--conservative
  super-Eddington accretion at $\sim10^{-6}\msunyr$. Conservative mass
  transfer begins as the mass ratio approaches unity and the accretion
  rate passes through the stable hydrogen burning band $\sim
  10^{-7}\msunyr$. The star appears as a supersoft X--ray binary
  during this phase, and the white dwarf accretes a few tenths of a
  solar mass. Once $q$ attains values $\la 1$ mass transfer proceeds
  on a nuclear timescale. Initially the accreted mass is lost in nova
  explosions so the white dwarf mass does not grow. Eventually the
  system reaches the regime discussed in this paper, where most
  transferred mass is burnt and retained by the WD during dwarf nova
  outbursts.  }
\label{Fig:TTMT}
\end{figure*}

We have seen from the previous Section that long--period dwarf novae
offer a route for making SNe Ia. Fig. \ref{Fig:progenitors} 
%%%%%%%%%%%%%%%%%%%%%%%%%%%%%%%%%%%%%
%%%%%%%%%%%%%%%%%%%%%%%%%%%%%%%%%%%%%
gives the regions of $M_{\rm 1,i}, M_{\rm 2,i}$ -- space where growth
to $M_{\rm Ch}$ may be possible. (This figure assumes conservative
mass transfer, but we will show in a future paper that similar results
are obtained when this assumption is relaxed.) The required $M_{\rm
1,i}$ is somewhat higher than typically results from single--star
evolution. This might seem to lead to the same need for massive
main--sequence progenitors and thus the same restriction on progenitor
numbers as for the SSS channel. However the nuclear--driven evolution
discussed here is of course also open to the outcomes of
thermal--timescale mass transfer, i.e. the SSS systems we discussed
earlier. Many of these, starting from typical white dwarf masses $\sim
0.7\msun$, have already increased them to values of the order required
in Fig. \ref{Fig:progenitors}. We therefore envisage a two--stage
process in the growth of $M_1$ towards $M_{\rm Ch}$
(Fig. \ref{Fig:TTMT}).

First, a SSS system with $M_1 \sim 0.7\msun$ and $M_2 > M_1$ undergoes
thermal--timescale mass transfer and increases the white dwarf mass to
values $\sim 1\msun$. Thermal--timescale mass transfer stops once the
mass ratio $q$ drops below a value $\sim 1$. What happens next to the
binary depends on the competition between orbital angular momentum
loss (presumably through magnetic stellar wind braking) and nuclear
expansion of the secondary (cf Schenker et al., 2002). If the angular
momentum loss timescale is shorter than that for nuclear evolution the
system becomes a short--period CV, often with some signs of chemical
evolution as in AE Aquarii (Schenker et al. 2002). If instead the
nuclear evolution timescale is shorter, the system will expand and
eventually become the kind of long--period CV we discuss here. For
suitable masses $M_{\rm 1,i}, M_{\rm 2,i}$ and orbital separation at
the end of thermal--timescale mass transfer the white dwarf mass can
grow to $M_{\rm Ch}$ and produce a SN Ia.

Assuming mass transfer starts at or near the end of the main sequence
and that the thermal--timescale phase has little effect on the core
mass of the donor when it reaches the base of the RGB, the minimum
donor mass which reaches the RGB with core mass $\gtsimeq 0.2\msun$ is
$\sim 2\msun$. For massive donors (mass ratios $q\gtsimeq$3--4)
sustained thermal--timescale mass transfer leads to dynamical
instability (Hjellming 1989; Kalogera \& Webbink 1996; Kolb et al.
2000).  This \emph{delayed dynamical instability} limits the maximum
donor mass before the thermal--timescale phase to $\sim3\msun$. Thus
the donor mass at the start of mass transfer is limited to the
approximate range 2 -- 3$\msun$.  Future work involving accurate
numerical modelling of the thermal--timescale evolution will refine
this simple picture.

\section{Conclusions}

We have shown that long--period dwarf novae offer a promising channel
for making Type Ia supernovae. In particular the white dwarf mass can
be raised to the Chandrasekhar value without the need to invoke fairly
massive and thus rare main--sequence progenitors. Moreover there is
very little danger of hydrogen contamination of the supernova ejecta.

Clearly more work is needed. In particular we need to calculate the
birthrate of such systems and compare this with the inferred rate
$\sim 3\times 10^{-3}\ {\rm yr}^{-1}$ of SNe Ia per galaxy. This in
turn requires an extensive grid of models. We shall return to this
problem in a future paper.
%%%%%%%%%%%%%%%%%%%%%%%%%%%%%%%%%%%%%
%%%%%%%%%%%%%%%%%%%%%%%%%%%%%%%%%%%%%

\section{Acknowledgments} 

Theoretical astrophysics research at Leicester is supported by a PPARC
rolling grant. ARK gratefully acknowledges a Royal Society Wolfson
Research Merit Award. We thank the referee, David Branch, for helpful
comments.

\end{document}